\documentclass[twocolumn,secnumarabic,amssymb,nobibnotes,aps,prl,nofootinbib]{revtex4-2}

\usepackage{graphicx, url}
\usepackage{color}
\usepackage{amsmath, amssymb,bbm}
\usepackage{epsfig}
\usepackage{rotating}
\usepackage{dcolumn}
\usepackage{bm}
\newcommand{\comment}[1]{}
\def \non{\nonumber}
\def \ra{\rightarrow}
\def \bea{\begin{eqnarray}}
\def \eea{\end{eqnarray}}
\def \Qbar{\overline Q}
\def \qqbar{Q\Qbar}

\def \jpsi{J/\psi}
\def \ccbar{c\overline{c}}

\def \sg{\sigma}

\begin{document}   
\title{Understanding $J/\psi$ and $\psi '$ production using a modified version of Non-Relativistic Quantum
Chromodynamics} 
\author{Sudhansu~S.~Biswal$^1$}
\email{sudhansu.biswal@gmail.com}
\author{Sushree~S.~Mishra$^1$}
\email{sushreesimran.mishra97@gmail.com}
\author{K.~Sridhar$^2$}
\email{sridhar.k@apu.edu.in}
\affiliation{$^1$ Department of Physics, Ravenshaw University,
Cuttack 753003, India\\
$^2$ School of Arts and Sciences, Azim Premji University, Sarjapura, Bangalore, 562125, India }

\begin{abstract}
\noindent There is serious disagreement between the predictions of Non-Relativistic Quantum 
Chromodynamics (NRQCD) and the data on $J/\psi$ polarisation which has persisted for almost 
a quarter of a century. We find that if we account for the effect of perturbative soft gluons
on the intermediate charm-anticharm octet states in NRQCD then the polarisation problem can be
resolved. In addition, this model, when used to fit the Run 1 data on $J/\psi$ and $\psi '$ 
production from the CDF experiment at Tevatron, gives good fits and yields values of (energy-
independent) non-perturbative parameters. These, in turn, can be used to make {\it parameter-free
predictions} for $J/\psi$ and $\psi '$ data from the CMS experiment at the Large Hadron 
Collider and the predictions are in excellent agreement with the CMS data.
\end{abstract}
\maketitle

\noindent When Non-Relativistic Quantum Chromodynamics (NRQCD)~\cite{bbl} was first
developed as an effective theory over twenty-five years ago, there was much hope that
it would shed light on understanding quarkonium production in much the same manner that
it had been already used to clarify the situation with respect to the decay of $P$-wave
quarkonia. It was, indeed, the most opportune time to explore the consequences of this
effective theory because the data from Tevatron \cite{cdf} on $J/\psi$ production had become available
and showed huge disagreement with the theoretical model that preceded NRQCD -- the 
Colour-Singlet Model (CSM)~\cite{berjon,br}. The success of NRQCD in bridging this
massive gap between theoretical predictions and experimental data went a long way in
furthering the belief of it being the correct theory of quarkonium production. But 
one could also discern a certain tendency among the community of
NRQCD theorists of regarding it as being something of a gospel truth -- the platitudinous use
of the word rigorous in NRQCD papers being a marker of this tendency -- but somewhere
in its claims to rigour lay the seeds of its nemesis.

That was to come from the prediction of the polarisation of the produced $J/\psi$ --
from the observation that a large fraction of them produced at large transverse
momentum, $p_T$, were from a fragmentation-like process -- with a single large-$p_T$
gluon fragmenting into an octet state (a ${}^3 S_1^8$ state, to be precise). The
appearance of the colour-octet component in the gluon fragmentation function and
its importance in understanding the data from the CDF experiment at the Tevatron 
was already explored \cite{jpsi} before the full understanding of the production 
mechanism in NRQCD was available. But, with the more complete understanding came
the realisation that, in this fragmentation-like process, the ${}^3 S_1^8$ $Q \bar Q$ 
state inherits the entire transverse polarisation of the (almost-real) gluon that
fragments into it and, subsequently, when the $Q \bar Q$ state evolves non-perturbatively
into a $J/\psi$, in the manner specified by NRQCD, the heavy-quark symmetry of NRQCD
ensures that almost all of this polarisation is bequeathed to the $J/\psi$ which, 
therefore, emerges transversely polarised at large-$p_T$ \cite{polar1}. This is true 
upto order $v^2$, where $v$ is the relative velocity between the $Q$ and the $\bar Q$ 
inside the bound state. This argument fails to hold at low or moderate $p_T$ where one 
expects the $J/\psi$ to be unpolarised and so the $p_T$ dependence of the polarisation, 
calculated in full detail in Ref. \cite{polar2}, is a crucial test of the theory. The theory 
failed the test miserably when data from the CDF experiment \cite{polar3} showed that the
polarisation of the $J/\psi$ was consistent with zero over all $p_T$.

It is useful to recall the basic ideas of NRQCD in somewhat more detail.
NRQCD is derived from the QCD Lagrangian by neglecting all states of momenta much larger 
than the heavy quark mass, $M_Q$ and to account for this exclusion by adding new interaction 
terms yielding the effective Lagrangian. The quarkonium state admits of a Fock-state
expansion in orders of $v$. At leading order, the $\qqbar$ state is in a colour-singlet
state but at {\cal O}$(v)$, it can be in a colour-octet state and connected to the
physical $J/\psi$ state through a non-perturbative gluon emission.  

The Fock space expansion of the physical $\jpsi$, which is a $^3S_1$
($J^{PC}=1^{--}$) state, is:
\bea
\left|\jpsi\right>={\cal O}(1)
        \,\left|\qqbar[^3S_1^{[1]}] \right>+
         {\cal O}(v^2)\,\left|\qqbar[^3P_J^{[8]}]\,g \right>+ \cr
        {\cal O}(v^4)\,\left|\qqbar[^1S_0^{[8]}]\,g \right>+
        {\cal O}(v^4)\,\left|\qqbar[^3S_1^{[8]}]\,gg \right>+
           \cdots 
\label{fockexpn}
\eea
In the above expansion the colour-singlet $^3S_1$ state contributes
at ${\cal O}(1)$. The $^3P_J$ contribution includes the three $P$ 
states ($J=0,\ 1,\ 2$). As the $P$-state production is itself down
by factor of ${\cal O}(v^2)$ both the colour-octet $P$ and $S$ channels
channels effectively contribute at the same order. The colour-octet 
state $^3P_J^{[8]}$ ($^1S_0^{[8]}$) becomes a physical $\jpsi$ by 
emitting a gluon in an E1 (M1) transition, while there is also a 
contribution at the same order from a $^3S_1^{[8]}$ state doing a 
double E1 transition to a $\jpsi$ state. 

The cross section for production of a quarkonium state $H$ can be 
factorised as:
\bea
  \sigma(H)\;=\;\sum_{n=\{\alpha,S,L,J\}} {F_n\over {M_Q}^{d_n-4}}
       \langle{\cal O}^H_n({}^{2S+1}L_J)\rangle, 
\label{factorizn}
\eea
where $F_n$'s are the short-distance coefficients and ${\cal O}_n$ are 
operators of naive dimension $d_n$, describing the long-distance effects.  
These non-perturbative matrix elements are guaranteed to be
energy-independent due to the NRQCD factorization formula, so that they
may be extracted at a given energy and used to predict quarkonium cross-sections
at other energies.

Writing this down explicitly in terms of the various octet and singlet intermediate states gives:
\begin{eqnarray}
\sigma_{J/\psi}  = \hat F_{{}^{3}S_1^{[1]}} \times \langle {\cal O} ({}^{3}S_1^{[1]}) \rangle +
                \hat F_{{}^{3}S_1^{[8]}} \times \langle {\cal O} ({}^{3}S_1^{[8]}) \rangle +\cr
                 \hat F_{{}^{1}S_0^{[8]}} \times \langle {\cal O} ({}^{1}S_0^{[8]}) \rangle 
                + {1 \over M^2} \biggl\lbrack\hat F_{{}^{3}P_J^{[8]}} \times \langle {\cal O} 
                     ({}^{3}P_J^{[8]}) \rangle \biggr\rbrack .
\label{Fock}
\end{eqnarray}

In Refs. \cite{cho1,cho2} the complete set of short-distance 
coefficients in NRQCD needed to study $J/\psi$ and $\chi$ production
was calculated and compared with the data from Tevatron \footnote{See also 
Ref. \cite{cgmp}. For a detailed review of quarkonium production
see Ref. \cite{brambilla}.}. These NRQCD calculations gave a good description
of the shapes of the $p_{_T}$ distributions of the charmonium resonances
at the Tevatron but the normalization of
these distributions was not predicted in NRQCD i.e. the non-perturbative
matrix elements which determined the normalization had to be obtained
by a fit to the data. Independent tests of the effective theory approach
were, therefore, necessary to determine the validity of the approach
and, indeed, other than the polarisation of the $J/\psi$ various proposals 
were made \cite{tests} to test NRQCD. But of all these, polarisation was
the litmus test of the theory and the serious disagreement of
NRQCD predictions with data on $J/\psi$ polarisation calls for a
critical evaluation of the theory.

It could be that the reason that the polarisation predictions are going awry is
that the charm quark is too light to be treated in NRQCD. However, if that is
the reason one cannot understand why the cross-sections for the charmonium resonances
work out right in NRQCD. The other line of attack has been to first note that
the colour-singlet model predicts zero polarisation and to then attempt
to jack up the colour-singlet contribution by invoking higher-order effects
in the singlet channel \cite{gong, artoisenet}. 
\footnote{For reviews of the status of these calculations and their experimental 
consequences, see Refs.~\cite{lansberg1, lansberg2}.} From the point of view
of the effective theory, it makes little sense to leave out one set of operators
(colour-octet) and work with only the other set (colour-singlet) without any
argument for the smallness of the octet operators and their effects in charmonium
production. The next-to-leading order corrections for the NRQCD matrix elements
have been computed \cite{higherorder} but then that does little to suggest a way
out of the polarisation problem. 

In the colour-singlet model, the $c \bar c$ pair is produced in the colour-singlet
state with the appropriate angular momentum and spin assignments and this
forms the physical quarkonium state through a non-perturbative transition. The description 
of the bound-state in NRQCD and, indeed, the computational procedure to obtain the
short-distance coefficients $F_n$ is the same as in the colour-singlet model: the
$c \bar c$ pair could be in a colour-singlet or octet state and, if it is in an
octet state in emits one or more {\it non-perturbative} gluons to make a transition
to the physical quarkonium state. The fact that a fixed number (one or two, in practice)
of gluons mediate this transition it is possible to figure out what the quantum numbers
of the octet $c \bar c$ would be given the quantum numbers of the quarkonium state.
In our proposed modification of the picture, this is the point that we wish to deviate 
from the usual narrative. The colour-octet $c \bar c$ state can radiate several 
soft {\it perturbative} gluons -- each emission taking away little energy but carrying 
away units of angular momentum. In the multiple emissions that the colour-octet state
can make before it makes the final NRQCD transition to a quarkonium state, the
angular momentum and spin assignments of the $c \bar c$ state changes constantly.
Perturbative soft gluon emission from colour-octet states to address the polarisation
problem has also been used in Ref. \cite{baranov} but it differs considerably from
the approach we present here. We also present a modified cross-section formula
with which we are able to fit the Tevatron cross-sections and present excellent
predictions for the LHC cross-sections measurements.

Let us consider $J/\psi$ production, to be specific. When an octet state transforms into 
a $J/\psi$ that process happens in NRQCD via a non-perturbative gluon and is subject to 
the counting rules and symmetries of NRQCD. The gluons that we are invoking are perturbative 
soft gluons that affect the intermediate perturbative state between the short-distance process 
and the final NRQCD transition. If one uses the NRQCD rules to pin down what octet states
can transform into a $J/\psi$ and also neglect the short-distance production of higher angular 
momentum states then the only octet states at the short-distance level that 
we need to consider are ${}^{3}S_1^{[8]}$, ${}^{3}P_{0,1,2}^{[8]}$, ${}^{1}S_0^{[8]}$ 
and ${}^1P_1^{[8]}$. If we label these states as $S_i$, $i=1, ..., 6$ then the
soft gluon emission process can be thought of as a stochastic process that mediates
transitions between these several states $S_i$. 

The resummation of leading logarithms resulting from the soft-gluon emission will give
rise to the familiar Sudakov form factor but the angular momentum of the radiating
object is left unchanged at this level. However, higher-order power-suppressed terms change this
picture -- the amplitude when expanded in powers of the threshold parameter yields
the angular-momentum operator and such terms can change the polarisation of the radiating
coloured particle \cite{low, magnea, ravindran}. 

In particular, if at large-$p_T$ we
were to produce a $^3S_1^{[8]} \equiv S_1$ state in a short-distance fragmentation-like process, this state
will now emit several soft gluons which will obliterate the transverse
polarisation that the state was produced with because the intermediate states $S_i$
produced in the process of soft-gluon emission will randomly oscillate between transverse
and longitudinal polarisation, yielding a net polarisation which is zero. The perturbative
soft gluons, thus, can give the right prediction for the polarisation of the $J/\psi$. We
do not attempt to calculate the dynamics of the soft-gluon emission explicitly -- but, indeed,
in the approach we follow we do not need to. It is sufficient to know that with each emission
the angular momentum of the coloured state can change. What we compute explicitly are the
cross-sections and for that we do not need a detailed understanding of the soft dynamics,
as we show below.

The price that has to be paid for bring in the perturbative soft gluons, 
however, is that we seem to lose the ability to
pin down the quantum numbers of the intermediate $c \bar c$ state. In the usual picture,
this was well under control: the state produced in the perturbative process was the one
that transitioned into the $J/\psi$ finally. With the soft gluons brought in, this is
no longer true. But we will see that, in spite of this apparent problem, we have enough
information to be able to compute the cross-sections. Because of the stochastic mixing
between the states $S_i$, the cross-section formula in Eqn. \ref{Fock} no longer holds.
If we assume our process is Markovian and the states $S_i$ freely mix with all the
transition probabilities being equal then we can write down a cross-section formula
as follows:
\begin{eqnarray}
\sigma_{J/\psi} &=& \biggl\lbrack \hat F_{{}^{3}S_1^{[1]}} 
                \times \langle {\cal O} ({}^{3}S_1^{[1]}) \rangle \biggr\rbrack \cr 
                &+& \biggl\lbrack  
                  \hat F_{{}^{3}S_1^{[8]}} 
                 + \hat F_{{}^{1}P_1^{[8]}} 
                + \hat F_{{}^{1}S_0^{[8]}} + (\hat F_{{}^{3}P_J^{[8]}} ) \biggr\rbrack 
                \times ({\langle {\cal O} ({}^{3}S_1^{[1]}) \rangle \over 8}) \cr
                &+& \biggl\lbrack  
                  \hat F_{{}^{3}S_1^{[8]}} 
                 + \hat F_{{}^{1}P_1^{[8]}} 
                + \hat F_{{}^{1}S_0^{[8]}} + (\hat F_{{}^{3}P_J^{[8]}} ) \biggr\rbrack 
                \times \langle {\cal O}  \rangle ,
\end{eqnarray}
where
\begin{equation}
     \langle {\cal O}  \rangle =
                \times \biggl\lbrack 
                 \langle {\cal O} ({}^{3}S_1^{[8]}) \rangle 
                + \langle {\cal O} ({}^{1}S_0^{[8]}) \rangle 
                + {\langle {\cal O} ({}^{3}P_J^{[8]}) \rangle \over M^2}
                    \biggr\rbrack 
\end{equation}

In contrast to the usual case, where we needed to fix three non-perturbative
parameters to get the $J/\psi$ cross-section, in our case it is the sum of these
parameters: so we have a single parameter to fit. 

To test the model that we have proposed, we start by using this cross-section formula
to make a fit to the data on $J/\psi$ and $\psi '$ production from the CDF experiment
at Tevatron \cite{cdf}. The $p_T$ distribution is given by the standard formula:
\bea
&&\frac{d\sg}{dp_{_T}} \;(p \bar p \ra \ccbar\; [^{2S+1}L^{[1,8]}_J]\, X)= \non \\
&&\sum \int \!dy \int \! dx_1 ~x_1\:G_{a/p} (x_1)~x_2\:G_{b/p}(x_2) 
\:\frac{4p_{_T}}{2x_1-\overline{x}_T\:e^y}\non\\
&&\frac{d\hat{\sg}}{d\hat{t}}
(ab\ra \ccbar[^{2S+1}L_J^{[1,8]}]\;d),
\label{eq:diff}
\eea
where the summation is over the partons ($a$ and $b$),    
the final state $\qqbar$ is in the $^1 S^{[1]}_0$, $^1 P^{[8]}_1$,  
$^3 S^{[8]}_1$ states and $G_{a/p}$,   
$G_{b/p}$ are the distributions of partons $a$ and $b$ in the
protons and $x_1$, $x_2$ are the respective  momentum they carry.
In the above formula, $\overline{x}_T=\sqrt{x_T^2+4\tau} \equiv 2 M_T/\sqrt{s}$ \ with \  
$x_T=2p_{_T}/\sqrt{s}$ and  \(\tau=M^2/s\).
$\sqrt{s}$ is the center-of-mass energy, 
$M$ is the mass of the resonance and $y$ is the rapidity at which the 
resonance is produced. 
The matrix  elements for the subprocesses 
corresponding to $F_1[^1S_0]$, $F_8[^3P_J]$ and $F_8[^3S_1]$ are listed in 
Refs.~\cite{cho2,gtw}. The remaining coefficient $F_8[^1P_1]$ has been 
calculated in~\cite{Mathews}.  

We assume, and it is reasonable to do so, that the $p_T$ distributions
of the final $J/\psi$ is not significantly different compared to that
of the octet state produced at the short-distance level. The multiple
gluon emission would at best change the normalisation of the cross-section
somewhat. We are fitting the normalisation, in any case, by comparing our
theoretical predictions to the data from Tevatron and so the effect of
the soft-gluons is accounted for in the fit.
The fits to the 1.8 TeV Tevatron data on $J/\psi$ and 
$\psi '$ production are shown in Fig. 1. Good fits to both sets of data
are obtained. 
\begin{figure}[h!]
\includegraphics[width=8cm]{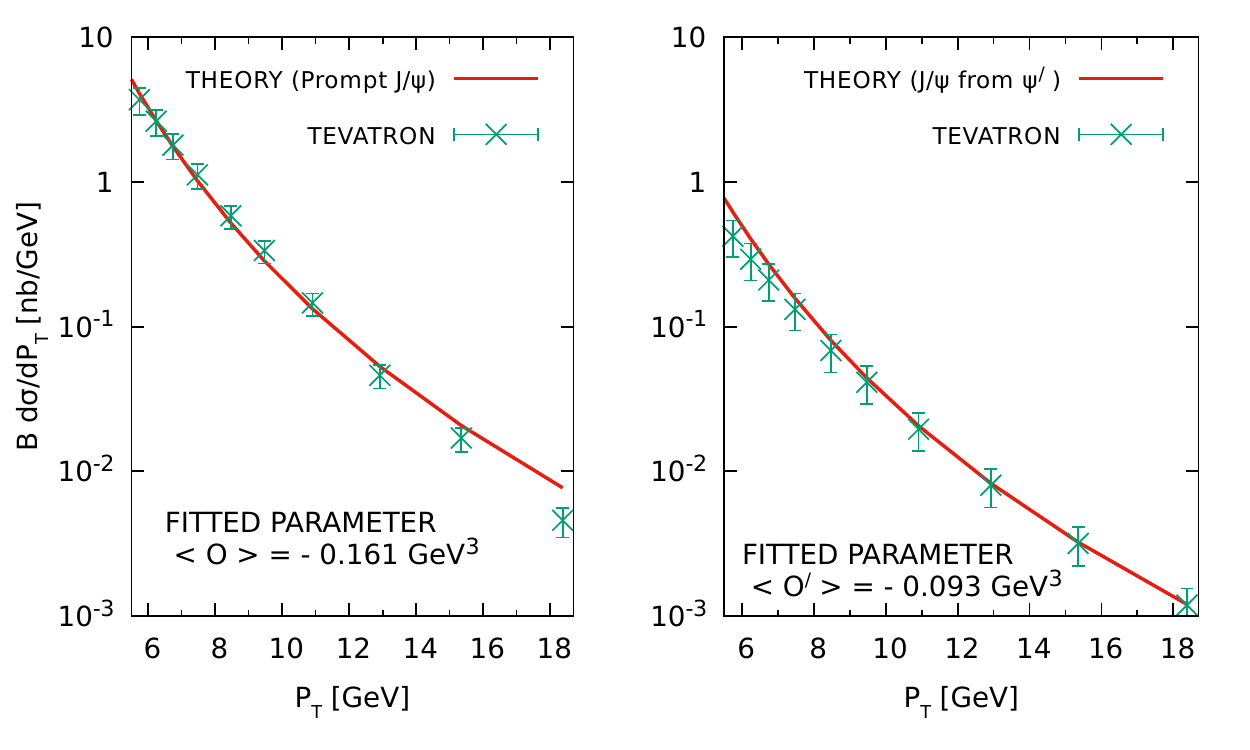}
\caption{Theoretical prediction of differential cross sections
fitted to the data on $J/\psi$ and $\psi '$ production from the CDF experiment
at Tevatron. }
\label{fig:1}
\end{figure}

The true test of our model is to see what predictions are obtained for
$J/\psi$ and $\psi '$ production at the Large Hadron collider runs of
13 TeV. We use the fitted non-perturbative parameters (one each for
$J/\psi$ and $\psi '$) to predict the distributions (in $p_T$ and in
bins of $y$) and compare them with data from the CMS experiment \cite{CMS:2017dju}.
As can be seen from, Fig. 2, the predictions compare very well with
the CMS data.
\begin{figure}[h!]
\includegraphics[width=8cm,height=8cm]{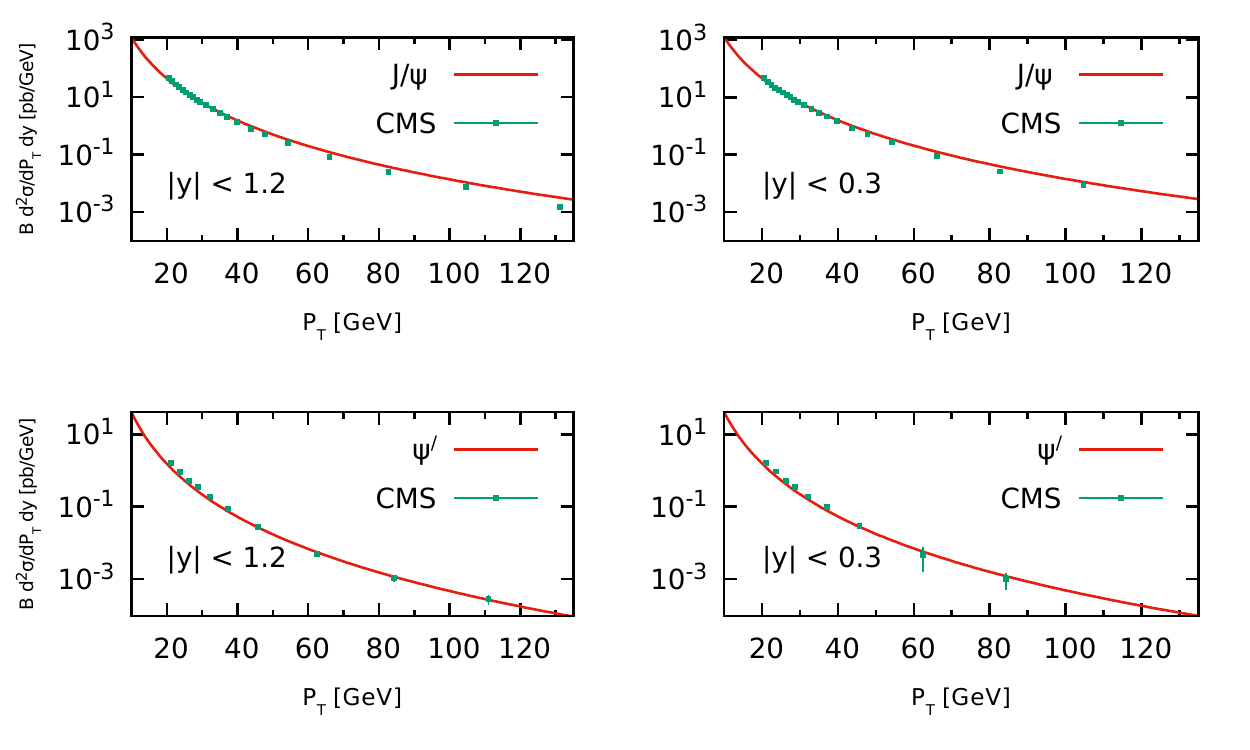}
\caption{ Predicted differential distributions for
$J/\psi$ and $\psi '$ production at the LHC running at
13 TeV compared  with data
from the CMS experiment.}
\label{fig:2}
\end{figure}

In conclusion, provoked by the long-standing disagreement of the predictions
of Non-Relativistic QCD (NRQCD) with the data on $J/\psi$ polarisation, we
have critically examined the theory and found it justified to study the
effect, neglected hitherto, of soft-gluon with the colour-octet $c \bar c$
pair that eventually forms a charmonium state like $J/\psi$. In complete
contrast to the usual picture, our model predicts that the produced $J/\psi$
is unpolarised which is in agreement with the polarisation data from Tevatron.
We have then fitted our model predictions to Tevatron data on $J/\psi$ and
$\psi '$ production and used the fitted parameters to {\it predict} the 
distributions at the LHC energy, and find excellent agreement with data
from the CMS experiment. 

The results are very encouraging and more studies of this model to understand
the production of other charmonium resonances and in other experimental
situations are being planned.
%

\section*{Acknowledgments}
The authors would like to acknowledge discussions with L. Magnea
and V. Ravindran and, in particular, R. Sharma, who was also involved
in the early stages of this project.


\end{document}